# Reconciling Spacetime and the Quantum: Relational Blockworld and the Quantum Liar Paradox

*W.M. Stuckey[1], Michael Silberstein[2,3] and Michael Cifone[3]*

## Abstract

The Relational Blockworld (RBW) interpretation of non-relativistic quantum mechanics (NRQM) is introduced. Accordingly, the spacetime of NRQM is a relational, non-separable blockworld whereby spatial distance is only defined between interacting trans-temporal objects. RBW is shown to provide a novel statistical interpretation of the wavefunction that deflates the measurement problem, as well as a geometric account of quantum entanglement and non-separability that satisfies locality per special relativity and is free of interpretative mystery. We present RBW's acausal and adynamical resolution of the so-called "quantum liar paradox," an experimental set-up alleged to be problematic for a spacetime conception of reality, and conclude by speculating on RBW's implications for quantum gravity.

[1] Department of Physics, Elizabethtown College, Elizabethtown, PA 17022, stuckeym@etown.edu
[2] Department of Philosophy, Elizabethtown College, Elizabethtown, PA 17022, silbermd@etown.edu
[3] Department of Philosophy, University of Maryland, College Park, MD 20742, cifonemc@wam.umd.edu

# 1. INTRODUCTION

Many philosophers and physicists expect to find new physics lurking in the answer to van Fraassen's[1] foundational question par excellence: *"how could the world possibly be the way quantum theory says it is?"* In fact, Smolin[2] believes that what "we are all missing" in the search for quantum gravity "involves two things: the foundations of quantum mechanics and the nature of time." We share this sentiment and are therefore motivated to "understand" non-relativistic quantum mechanics (NRQM). As we will show, our interpretation has strong implications for the practice and unification of physics, and we will speculate formally on these consequences.

Since there are several well-known conceptual and formal tensions between relativity and quantum mechanics which bear on the project of unifying general relativity (GR) and quantum field theory (QFT), we feel that a necessary condition for "understanding" NRQM is to couch it in space and time as required for "comprehension" per Schrödinger[3],

> This contradiction is so strongly felt that it has even been doubted whether what goes on in an atom can be described within the scheme of space and time. From a philosophical standpoint, I should consider a conclusive decision in this sense as equivalent to a complete surrender. For we cannot really avoid our thinking in terms of space and time, and what we cannot comprehend within it, we cannot comprehend at all.

and Einstein[4],

> Some physicists, among them myself, cannot believe that we must abandon, actually and forever, the idea of direct representation of physical reality in space and time.

As Howard notes in the following passage, one of the central debates between the founding fathers of quantum mechanics was over the conflict between the spacetime picture and the quantum picture of reality and how they may be reconciled[5]:

> The second striking feature of Pauli's last-quoted paragraph is that it points backward to what was by 1935 an *old* debate over the nonseparable manner in which quantum mechanics describes interacting systems. The fact that this was the central issue in the pre-1935 debate over the adequacy of the quantum theory disappeared from the collective memory of the physics community after EPR….Einstein had been trying in every which way to convince his colleagues that this was sufficient reason to abandon the quantum path…But it was not just

Einstein who worried about quantum nonseparability in the years before 1935. It was at the forefront of the thinking of Bohr and Schrödinger.

In today's terminology we would say that the spacetime picture of relativity adheres to the following principles[6]:

Separability principle: any two systems A and B, regardless of the history of their interactions, separated by a non-null spatiotemporal interval have their own independent real states such that the joint state is completely determined by the independent states.

Locality principle: any two space-like separated systems A and B are such that the separate real state of A let us say, cannot be influenced by events in the neighborhood of B.

It is now generally believed that Einstein-Podolsky-Rosen (EPR) correlations, i.e., correlated space-like separated experimental outcomes which violate Bell's inequality, force us to abandon either the separability or locality principle.

As Howard notes, Einstein thought that both these principles, but especially the latter, were transcendental grounds for the very possibility of science. Einstein's spatiotemporal realism is summarized in his own words[7]:

Is there not an experiential reality that one encounters directly and that is also, indirectly, the source of that which science designates as real? Moreover, are the realists and, with them, all natural scientists not right if they allow themselves to be led by the startling possibility of ordering all experience in a (spatio-temporal-causal) conceptual system to postulate something real that exists independently of their own thought and being?

Minkowski spacetime (M4) is a perfect realization of Einstein's vision but as Howard says[8]:

Schrödinger's introduction of entangled n-particle wave functions written not in 3-space but in 3n-dimensional configuration space offends against space-time description because it denies the mutual independence of spatially separated systems that is a fundamental feature of a space-time description.

And we might add that realism about configuration space also destroys Einstein's vision of spacetime as the be-all and end-all of reality as exemplified by M4.

All of this raises an interesting question about just how much of the spacetime picture can be retained given quantum mechanics. As we will show, the Relational

Blockworld[9] interpretation of NRQM points to a far more intimate and unifying connection between spacetime and the quantum than most have appreciated. Many will assume that such a geometric interpretation is impossible because quantum wavefunctions live in Hilbert space and contain much more information than can be represented in a classical space of three dimensions. As Peter Lewis says[10], "the inescapable conclusion for the wavefunction realist seems to be that the world has $3N$ dimensions; and the immediate problem this raises is explaining how this conclusion is consistent with our experience of a three-dimensional world." On the contrary, the existence of the non-commutativity of quantum mechanics is deeply related to the structure of *spacetime* itself, without having to invoke the geometry of Hilbert space. Specifically, as will be demonstrated in section 2, the non-commutativity of NRQM's position and momentum operators *is a consequence of* the relativity of simultaneity. Since, as will also be demonstrated in section 2, the NRQM density operator can be obtained from the spacetime symmetries of the experimental configuration, we justify a Relational Blockworld (RBW) interpretation of NRQM.

*1.1 Caveats.* It is important not to be misled at this early stage by our claim about the spacetime symmetries of the experimental configuration. We are not advocating a brute spatiotemporal relationalism between sources and detectors, themselves conceived as classical and substantial trans-temporal, macroscopic objects. Rather, it's "relations all the way down" to echo Mermin. The spacetime symmetries methodology of NRQM is just the beginning of our account wherein "it is all related" because "it is all relations." That is, on our view any given relatum (such as a source or detector) always turns out to be a relational structure itself upon further analysis. The formal characterization of relations will change accordingly as we move toward the more fundamental relations underlying RBW (as introduced in section 2), but at the level of experimental set-ups in NRQM, spacetime symmetries are the most appropriate characterization of relations (as illustrated in section 4). In short, relationalism does not end with macroscopic objects but applies to their ultimate "constituents" as well.

The reader is warned that RBW is counterintuitive in many respects. Of course there are many interpretations of quantum mechanics that have highly counterintuitive features, but RBW possesses its own unique twists on several such features. Primarily, these counter-intuitive aspects arise from: (1) our claim that relations are fundamental to relata and (2) our particular variation of the blockworld.

*1.2 Relations Fundamental to Relata.* Assuming relations are fundamental to relata is not unique to RBW. For example Carlo Rovelli's relational interpretation of quantum mechanics[11] holds that a system's states or the values of its physical quantities as standardly conceived only exist relative to a cut between a system and an observer or measuring instrument. As well, on Rovelli's relational account, the appearance of determinate observations from pure quantum superpositions happens only relative to the interaction of the system and observer. Rovelli is rejecting absolutely determinate relata. Rovelli's relational interpretation of NRQM is inspired by Einstein's theory of special relativity in two respects. First, he makes the following analogy with special relativity: relational quantum mechanics relativizes states and physical quantities to observers the way special relativity relativizes simultaneity to observers. Second, Einstein does not merely provide an interpretation of the Lorentz formalism, but he derives the formalism on the basis of some simple physical principles, namely the relativity principle and the light postulate[12].

Another closely related example is Mermin's Ithaca interpretation[13] which tries to "understand quantum mechanics in terms of statistical correlations without there being any determinate correlata that the statistical correlations characterize[14]." According to Mermin, physics, e.g., quantum mechanics, is about correlations and only correlations; "it's correlations all the way down." It is not about correlations between determinate physical records nor is it about correlations between determinate physical properties. Rather, physics is about correlations without correlata. On Mermin's view, correlations have physical reality and that which they correlate does not. Mermin claims that the physical reality of a system consists of the (internal) correlations among its subsystems and its (external) correlations with other systems, viewed together with itself as subsystems of a larger system. Mermin also claims inspiration from special relativity.

RBW shares with the relational and Ithaca interpretations a rejection of the notion of absolute states and properties. RBW also shares inspiration from relativity but as we shall see, RBW provides a much deeper and more unifying relationship between quantum mechanics and special relativity than the relational or Ithaca interpretations. In addition, both formally and conceptually, the characterization of relationalism in RBW is quite different than either of these views.

First, in terms of specific formalism, RBW employs spacetime symmetries and relations fundamental to those symmetries best characterized *as a* mathematical co-construction of things, space and time (explained in section 2). Second, the rubric characterizing relationalism is ontological structural realism[15] (OSR), which rejects the idea that reality is ultimately *composed of things,* i.e., self-subsisting entities, individuals or trans-temporal objects[16] with intrinsic properties and "primitive thisness," haecceity, etc. According to OSR the world has an objective modal *structure* that is ontologically fundamental, in the sense of not supervening on the intrinsic properties of a set of individuals. In Einstein's terminology, given OSR, particles do not have their own "being thus." The objective modal structure of the world and the abstract structural relations so characterized are fundamental features of reality relative to *entities* such as particles, atoms, etc. This is not anti-realism about objects or relata, but a denial of their fundamentality. Rather, relations are primary while the things are derivative, thus rejecting "building block" atomism or Lego-philosophy. Relata inherit their individuality and identity from the structure of relations. According to RBW, entities/objects and even the dynamical laws allegedly "governing things" are secondary to relational structure.

While the standard conception of structure is either set theoretic or logical, OSR holds that graph theory provides a better formal model for the nature of reality because relations are fundamental to nodes therein[17]. Many people have argued that giving primacy to relations and abstracting relata from them is somehow incoherent. However, graph theory shows us that such objections are prejudiced by atomistic thinking and ordinary language. In fact, per Leitgeb and Ladyman[18] the identity and diversity of individuals in a structure are primitive features of the structure as a whole in graph theory. Thus, we employ a spatiotemporal graph to provide a heuristic characterization of RBW in section 2.

What this implies for the quantum domain is that one must be cautious in using everyday classical metaphysics of individuality. For example, it is quite common for physicists to say things like, "I can see a glowing atom in the Pauli trap." RBW *a la* OSR does not deny such a claim so long as it is not meant to imply any "being thus" on the part of the atom, a metaphysical *interpretation* not entailed by the facts. Certainly, it is difficult to think about structure without "hypostatizing" individuals or relata as the bearers of structure, but it does not follow that relata are truly ontologically fundamental.

None of this is really new as OSR-type views have a long and distinguished history in foundational physics[19] and group theoretic accounts[20] of objects have a long history in the development of quantum mechanics. The group-theoretic conception of the 'constitution' of objects as sets of invariants under symmetry transformations can be found in the writings of Cassirer[21], Eddington[22], Schrödinger[23], Lyre[24], and Weyl[25]. When it comes to fundamental physics, objects are very often identified via group-theoretic structure, e.g., quantum field theory. So, while counterintuitive, the notion of relations being fundamental to relata is not without precedence.

*1.3 The Blockworld.* The second counterintuitive feature of RBW is the use of a blockworld (BW) in the explanation and interpretation of quantum mechanics. Thus, to appreciate the RBW ontology, one must appreciate the blockworld perspective[26], i.e.,

> There is no dynamics within space-time itself: nothing ever moves therein; nothing happens; nothing changes. In particular, one does not think of particles as moving through space-time, or as following along their world-lines. Rather, particles are just in space-time, once and for all, and the world-line represents, all at once, the complete life history of the particle.

When Geroch says that "there is no dynamics within space-time itself," he is not denying that the mosaic of the blockworld possesses patterns that can be described with dynamical laws. Nor is he denying the predictive and explanatory value of such laws. Rather, given the reality of all events in a blockworld, dynamics are not "event factories" that bring heretofore non-existent events (such as measurement outcomes) into being. Dynamical laws are not brute unexplained explainers that "produce" events. Geroch is advocating for what philosophers call Humeanism about laws. Namely, the claim is that dynamical laws are *descriptions of regularities* and not the *brute explanation* for such regularities. His point is that in a blockworld, Humeanism about laws is an obvious position to take because everything is just "there" from a "God's eye" (Archimedean) point of view. That is, all events past, present and future are equally "real" in a blockworld.

Others have suggested that we ought to take the fact of BW seriously when doing physics and modeling reality. For example, Huw Price[27] calls it the "Archimedean view from nowhen" and it has motivated him to take seriously the idea of a time-symmetric quantum mechanics and so-called backwards causation in quantum mechanics (BCQM). As he says about his book defending BCQM[28], "the aim of the book is to explore the consequences of the block universe view in physics and philosophy." Price is attempting to construct a local hidden-variables interpretation of NRQM that explains quantum non-locality with purely time-like dynamics or backwards causation. According to Price, BCQM provides an explanation of the Bell correlations[29] "which shows that they are not really non-local at all, in that they depend on purely *local* interactions between particles and measuring devices concerned. They *seem* non-local only if we overlook the present relevance of future interactions."

The key explanatory move that Price makes is to have information travel backwards along the light-cones of the two EPR particles, converging at the source of the entangled state. Presumably, this is the point in spacetime where the entangled state is "prepared." The picture we must think of is this: the future measurement interaction in separate wings of an EPR apparatus is the cause of the (earlier) entangled state, so the "point at which they separate" is the "effect" of a causal chain "originating" with the measurement interaction. This is to put the point directly in terms of *backwards* causation. The arrow of causation does not point from one spacelike separated wing of the apparatus to the other, across *space*, but rather it points backwards in *time* to the point at which the particles separated. Other blockworld motivated accounts of quantum mechanics include those by Cramer[30], Lewis[31] and Barrett[32].

The connection between BCQM or time-symmetric accounts of the quantum and the BW is straightforward: in a BW the state preparations and measurement outcomes are equally real, i.e., already "there." Thus, since a dynamic interpretation of the BW picture is superfluous, one might as well claim the measurement outcomes "effect the state preparations" rather than the converse. Of course it may seem trivial to explain the outcomes of quantum experiments (or anything else) using the BW. After all, one could answer *any* question in this vein by saying something like "it's all just there in the BW, end of story." In order to avoid trivializing the BW explanation, BW interpretations of

NRQM invoke clever devices such as time-like backwards causation[33], advanced action[34] and the two-vector formalism[35]. Do these beautiful and clever devices really avoid the charge of triviality? Such explanations are no less *dynamical* than standard quantum mechanics, which is puzzling given that the original blockworld motivation for such accounts lacks *absolute* change and becoming. As far we know, only Cramer speaks to this worry. Cramer notes that the backwards-causal elements of his theory are "only a pedagogical convention," and that in fact "the process is atemporal[36]." Indeed, it seems to us that all such dynamical or causal devices in a BW should be viewed fundamentally as book keeping. BCQM and the like, even having acknowledged the potential explanatory importance of BW, have not gone far enough in their atemporal, acausal and adynamical thinking. Whereas such accounts are willing to think backwardly, temporally speaking, it is still essentially *dynamical, temporal* thinking.

We rather believe the key to rendering a BW explanation nontrivial is to provide an algorithm for the relevant BW construction. Thus, the answer to "Why did X follow Y and Z?" is not merely, "Because X is already 'there' in the future of Y and Z per the BW," but as we will illustrate, "Because this must be the spatiotemporal relationship of X, Y and Z in the BW per the self-consistent definition of the entities involved in X, Y and Z." If one chooses to read dynamical stories from a BW picture, they may where feasible. However, BW descriptions are not limited to the depiction of dynamical/causal phenomena, so they are not constrained to dynamical/causal storytelling. In the following passage Dainton[37] paints a suggestive picture of what it means to take the BW perspective seriously both ontologically and explanatorily:

> Imagine that I am a God-like being who has decided to design and then create a logically consistent universe with laws of nature similar to those that obtain in our universe…Since the universe will be of the block-variety I will have to create it *as a whole*: the beginning, middle and end will come into being together…Well, assume that our universe is a static block, even if it never 'came into being', it nonetheless exists (timelessly) as a coherent whole, containing a globally consistent spread of events. At the weakest level, "consistency" here simply means that the laws of logic are obeyed, but in the case of universes like our own, where there are universe-wide laws of nature, the consistency constraint is stronger: everything that happens is in accord with the laws of nature. In saying that the consistency is "global" I mean that the different parts of the universe all have to fit smoothly together, rather like the pieces of a well-made mosaic or jigsaw puzzle.

Does reality contain phenomena which *strongly suggest* an acausal BW algorithm? According to RBW, the deepest explanation of EPR-Bell correlations is such an algorithm. NRQM *a la* RBW provides an acausal BW algorithm in its prediction of Bell inequality violations and these violations have been observed. So it appears that reality does harbor acausal BW phenomena and NRQM *a la* RBW is one algorithm for depicting the self-consistent placement of such phenomena in a blockworld, as will be illustrated via the quantum liar experiment in section 4.

We support this claim in section 2 by first reviewing a result in which the non-commutativity of NRQM's position and momentum operators is a consequence of the relativity of simultaneity, and as is well known the latter implies a blockworld barring some neo-Lorentzian adornment, re-interpretation or the like[38]. The second result reviewed in section 2 shows the density operator of an experimental configuration is obtained from the "past, present and future" of the entire spatiotemporal configuration *a la* the spacetime symmetries of the experimental set-up: from the initiation of the experiment to its outcomes (as is clear, for example, in the path-integral formalism). The blockworld as implied by the spacetime picture does real explanatory and unifying work in RBW. Thus RBW helps to unify the quantum and spacetime formally, conceptually and metaphysically in ways that neither other relational accounts nor BW-motivated accounts have to date. For all these reasons we claim that RBW constitutes a geometric, acausal and adynamical account of NRQM and spacetime that is fundamental to dynamical explanations. As Dainton says[39]:

> If this strikes us as odd it is because we are unused to thinking of the universe as a vast spatiotemporal mosaic, but if the universe *is* a vast spatiotemporal mosaic, then, given the reality of the future, the future determines the past as much as the past determines the future. The constraints that later events place on earlier ones are not always causal [or dynamical or in any way time-like]. It is more typically a matter of coordination: the future events exist in the same universe as the earlier events, in a coherent, smooth-fitting, law-abiding whole.

*1.4 Non-separability of Spacetime Regions and Quantum States*. The blockworld of RBW is precisely in keeping with Geroch's "all at once" notion of spacetime and Dainton's "vast spatiotemporal mosaic," but it is important to note that it is a non-separable BW while that of relativity theory is separable. That is to say, the metric field of relativity theory takes on values at each point of the differentiable spacetime manifold, even in regions where the stress-energy tensor is zero, as if "things" are distinct from the concepts of space and time. Per RBW, the concepts of space, time and trans-temporal objects can only be defined self-consistently so each is meaningless in the absence of the others. In section 2, we suggest a method to formalize this idea, deriving a spatial distance defined only between interacting trans-temporal objects. Accordingly, there need not be an 'exchange' particle or wave moving 'through space' between the worldlines of trans-temporal objects to dynamically mediate their interaction and establish their spatial separation. As a consequence, we understand that an NRQM detection event (subset of the detector) results from a particular, rarefied subset of the relations defining sources, detectors, beam splitters, mirrors, etc. in an "all at once" fashion. In this picture, there are no "screened off" particles moving in a wave-like fashion through separable elements of the experimental arrangement to cause detection events, but rather such detection events are evidence that the experimental equipment itself is non-separable[1]. While non-separable, RBW upholds locality in the sense that there is no action at a distance, no instantaneous dynamical or causal connection between space-like separated events. And, there are no space-like worldlines in RBW. Thus, we have *the non-separability of dynamical entities, e.g., sources and detectors, while the entities themselves respect locality*. In this sense, we agree with Howard[40] that NRQM is best understood as violating "separability" (i.e., independence) rather than "locality" (i.e., no action at a distance, no super-luminal signaling), and we take to heart Pauli's admonition that[41] "in providing a systematic foundation for quantum mechanics, one should start more from the composition and separation of systems than has until now (with Dirac, e.g.) been the case."

One might perceive a certain tension in the combination of relationalism and the BW perspective. After all, nothing seems more *absolute* than the BW viewed as a whole, hence the Archimedean metaphor. One can just imagine Newton's God gazing upon the timeless, static 4-dimensional BW mosaic (her sensorium) from her perch in the fifth (or higher) dimension; what could be more absolute? But relationalism is a rejection of the absolute and the very idea of a God's eye perspective. In any case, one must never forget that while RBW is a blockworld in the sense that all events are equally real, it is a *relational* blockworld so there is no meaning to a God's eye perspective, i.e., any beings observing the BW must be *a part of it*. Short of occupying all the perspectives "at once," there is nothing that corresponds to such a privileged view.

---

[1] Since space, time and trans-temporal objects are to be mutually and self-consistently defined (via relations), the non-separability of spacetime entails the non-separability of trans-temporal objects and vice-versa. RBW does away with any matter/geometry dualism.

*1.5 Paper Overview.* We offer a graphical model for this non-separable, relational blockworld in section 2. In support of our heuristic model, we introduce the formalism of RBW by outlining results due to Kaiser, Anandan, Bohr, Ulfbeck, and Mottelson, and speculating on a spatiotemporally discrete approach underlying NRQM and QFT. We propose this spatiotemporally discrete approach both to follow up on the consequences of RBW for the practice and unification of physics, and to illustrate the RBW ontology. In section 3, we use this relational, non-separable blockworld to provide a geometric account of quantum entanglement and non-separability that is free of conflict with the locality of SR and free of interpretative mystery. Therein, we also show how RBW provides a novel statistical interpretation of the wavefunction that deflates the measurement problem. To illustrate the nature of explanation for NRQM phenomena in a relational blockworld, we use RBW to resolve the so-called "quantum liar paradox" in section 4. Speculations on the possible implications for quantum gravity and the spacetime structure of GR are found in section 5.

## 2. THE RELATIONAL BLOCKWORLD

The RBW interpretation of NRQM is founded, in part, on a result due to Kaiser[42], Bohr & Ulfbeck[43] and Anandan[44] who showed independently that the non-commutivity of the position and momentum operators in NRQM follows from the non-commutivity of the Lorentz boosts and spatial translations in SR, i.e., the relativity of simultaneity. Whereas Bohr *et al.* maintain a dynamical view of NRQM via the Theory of Genuine Fortuitousness[2], we assume the blockworld implication of the relativity of simultaneity so that no particular event is more fortuitous than any other. Kaiser writes[45],

> For had we begun with Newtonian spacetime, we would have the Galilean group instead of [the restricted Poincaré group]. Since Galilean boosts

---
[2] As with RBW, detector clicks are not caused by impinging particles; in fact they're not caused by any*thing*, and NRQM simply provides the distributions of uncaused clicks. Since Bohr *et al.* do not further assume that the detector itself is a collection of fortuitous events, they seem to distinguish between a macroscopic, causal world and a microscopic fortuitous world.

> commute with spatial translations (time being absolute), the brackets between the corresponding generators vanish, hence no canonical commutation relations (CCR)! In the [c → ∞ limit of the Poincaré algebra], *the CCR are a remnant of relativistic invariance where, due to the nonabsolute nature of simultaneity, spatial translations do not commute with pure Lorentz transformations*. [Italics in original].

Bohr & Ulfbeck also realized that the "Galilean transformation in the weakly relativistic regime[46]" is needed to construct a position operator for NRQM, and this transformation "includes the departure from simultaneity, which is part of relativistic invariance." Specifically, they note that the commutator between a "weakly relativistic" boost and a spatial translation results in "a time displacement," which is crucial to the relativity of simultaneity. Thus they write[47],

> "For ourselves, an important point that had for long been an obstacle, was the realization that the position of a particle, which is a basic element of nonrelativistic quantum mechanics, requires the link between space and time of relativistic invariance."

So, *the essence of non-relativistic quantum mechanics – its canonical commutation relations – is entailed by the relativity of simultaneity*.

To outline Kaiser's result, we take the limit c → ∞ in the following bracket of the Lie algebra of the Poincaré group:

$$[T_m, K_n] = \frac{-i}{c^2} \delta_{mn} T_0 \qquad (1)$$

where subscripts m and n take values of 1, 2 and 3, $T_0$ is the generator of time translations, $T_m$ are the generators of spatial translations, $K_n$ are the boost generators, $i^2 = -1$, and c is the speed of light. We obtain

$$[T_m, K_n] = \frac{-i}{\hbar} \delta_{mn} M \qquad (2)$$

where M is obtained from the mass-squared operator in the c → ∞ limit since

$$c^{-2} \hbar T_0 = c^{-2} P_0 \qquad (3)$$

and

$$\frac{P_o}{c^2} = (M^2 + c^{-2} P^2)^{1/2} = M + \frac{P^2}{2Mc^2} + O(c^{-4}) \qquad (4).$$

Thus, $c^{-2}T_0 \to \dfrac{M}{\hbar}$ in the limit c → ∞. [M ≡ mI, where m is identified as "mass" by choice of 'scaling factor' ℏ.] So, letting

$$P_m \equiv \hbar T_m \tag{5}$$

and

$$Q_n \equiv \dfrac{\hbar}{m} K_n \tag{6}$$

we have

$$[P_m, Q_n] = \dfrac{\hbar^2}{m}[T_m, K_n] = \left(\dfrac{-\hbar^2}{m}\right)\left(\dfrac{i}{\hbar}\right)\delta_{mn} mI = -i\hbar \delta_{mn} I \tag{7}.$$

Bohr & Ulfbeck point out that in this "weakly relativistic regime" the coordinate transformations now look like:

$$\begin{aligned} X &= x - vt \\ T &= t - \dfrac{vx}{c^2} \end{aligned} \tag{8}.$$

These transformations differ from Lorentz transformations because they lack the factor

$$\gamma = \left(1 - \dfrac{v^2}{c^2}\right)^{-1/2} \tag{9}$$

which is responsible for time dilation and length contraction. And, these transformations differ from Galilean transformations by the temporal displacement vx/c² which is responsible for the relativity of simultaneity, i.e., in a Galilean transformation time is absolute so T = t. Therefore, the spacetime structure of Kaiser *et al.* (K4) lies between Galilean spacetime (G4) and M4, and we see that the Heisenberg commutation relations are not the result of Galilean invariance, where spatial translations commute with boosts, but rather they result from the relativity of simultaneity per Lorentz invariance.

      The received view has it that Schrödinger's equation is Galilean invariant, so it is generally understood that NRQM resides in G4 and therefore respects absolute simultaneity[48]. *Prima facie* the Kaiser *et al.* result seems incompatible with the received view, so to demonstrate that these results are indeed compatible, we now show that these results do not effect the Schrödinger dynamics[49]. To show this we simply operate on |ψ> first with the spatial translation operator then the boost operator and compare that

outcome to the reverse order of operations. The spatial translation (by *a*) and boost (by v) operators in x are

$$U_T = e^{-iaT_x} \quad \text{and} \quad U_K = e^{-ivK_x} \tag{10}$$

respectively. These yield

$$U_K U_T |\psi\rangle = U_T U_K e^{-iavmI/\hbar} |\psi\rangle \tag{11}.$$

Thus, we see that the geometric structure of Eq. 7 introduces a mere phase to |ψ> and is therefore without consequence in the computation of expectation values. And in fact, this phase is consistent with that under which the Schrödinger equation is shown to be Galilean invariant[50].

Therefore, we realize that the spacetime structure for NRQM, while not M4 in that it lacks time dilation, length contraction and separability, nonetheless contains a "footprint of relativity[51]," i.e., the relativity of simultaneity. In light of this result, it should be clear that there is no metaphysical tension between SR and NRQM. This formal result gives us motivation for believing that NRQM is intimately connected to the geometry of spacetime consistent with the relativity of simultaneity and therefore we feel justified in couching an interpretation of NRQM in a blockworld, albeit a non-separable blockworld in which relations are fundamental to relata.

That relations are fundamental to trans-temporal objects, as opposed to the converse per a dynamic perspective, can be justified via the work of Bohr, Mottelson & Ulfbeck[52] who showed how the quantum density operator can be obtained via the symmetry group of the relevant observable. Their result follows from two theorems due to Georgi[53], i.e.,

> The matrix elements of the unitary, irreducible representations of G are a complete orthonormal set for the vector space of the regular representation, or alternatively, for functions of g ∈ G.

which gives[54]

> If a hermitian operator, H, commutes with all the elements, D(g), of a representation of the group G, then you can choose the eigenstates of H to transform according to irreducible representations of G. If an irreducible representation appears only once in the Hilbert space, every state in the irreducible representation is an eigenstate of H with the same eigenvalue.

What we mean by "the symmetry group" is precisely that group G with which some observable H commutes (although, these symmetry elements may be identified without actually constructing H). Thus, the mean value of our hermitian operator H can be calculated using the density matrix obtained wholly by D(g) and <D(g)> for all g ∈ G. Observables such as H are simply 'along for the ride' so to speak.

While we do not reproduce Bohr *et al.*'s derivation of the density matrix, we do provide a prefacing link with Georgi's theorems. Starting with Eq. 1.68 of Georgi[55],

$$\sum_g \frac{n_a}{N} [D_a(g^{-1})]_{kj} [D_b(g)]_{lm} = \delta_{ab} \delta_{jl} \delta_{km} \qquad (12)$$

where $n_a$ is the dimensionality of the irreducible representation, $D_a$, and N is the group order, and considering but one particular irreducible representation, D, we obtain the starting point (orthogonality relation) found in Bohr *et al.* (their Eq. 1),

$$\sum_g \frac{n}{N} [D(g^{-1})]_{kj} [D(g)]_{lm} = \delta_{jl} \delta_{km} \qquad (13)$$

where n is the dimension of the irreducible representation. From this, they obtain the density matrix as a function of the irreducible representations of the symmetry group elements, D(g), and their averages, <D(g)>, i.e., (their Eq. 6):

$$\rho \equiv \frac{n}{N} \sum_g D(g^{-1}) \langle D(g) \rangle \qquad (14).$$

The methodological significance of the Bohr *et al.* result is that any NRQM system may be described with the appropriate *spacetime* symmetry group. The philosophical significance of this proof is more interesting, and one rooted in RBW's ontology of spacetime relationalism. This ontology, as we will argue in the following section, easily resolves the conceptual problems of NRQM while conveying an underlying unity between SR and NRQM.

Exactly what it means to say relations are fundamental to relata will be illustrated technically for NRQM by the example in section 4 in terms of the spacetime symmetries of the experimental configuration, and an even more fundamental conception of relationalism will be outlined via the proposed spatiotemporally discrete formalism in the remainder of this section, but we pause here to introduce the idea heuristically via a graphical representation of a non-separable blockworld. Figure 1 shows the links of a graph for two (implied) worldlines in a relational G4. The vertical links (temporal translations) are generated by the Hamiltonian and the horizontal links (spatial translations) are generated by the momentum. Since boosts commute with spatial translations, the boosted version looks the same, i.e., spatial hypersurfaces are the same for observers in relative motion. Therefore, the only way to move along worldline 1 or 2 is via vertical links, i.e., horizontal displacement between worldlines cannot result in any temporal displacement along the worldlines. This represents the temporal Galilean transformation, T = t, consistent with presentism. In a spacetime where boosts don't commute with spatial translations, the temporal coordinate transformation contains a translation, e.g., vx/c$^2$ in Eq. 8. A relational spacetime of this type is represented graphically in Figure 2. In this type of spacetime it is possible to move along worldline 1 or 2 temporally by moving between the worldlines using the boosted spatial hypersurfaces, thus the blockworld implication. If spatial distance is only defined via the horizontal links between worldlines, then we say the spacetime is *non-separable* as explained in section 1.4.

In an effort to formalize the idea that spatial separation exists only between interacting trans-temporal objects[56], we are exploring a spatiotemporally discrete formalism underlying quantum physics with NRQM following in the spatially discrete, temporally continuous limit and QFT following in the limit of both spatial and temporal continuity (Figure 3). This approach constitutes a *unification* of physics as opposed to a mere *discrete approximation* thereto, since we are proposing a source for the action, which is otherwise fundamental. So, for example, the spatiotemporally discrete counterpart to the QFT transition amplitude for interacting sources without scattering

$$Z = \int D\varphi \exp\left[i\int d^4x \left[\frac{1}{2}(d\varphi)^2 - V(\varphi) + J(x)\varphi(x)\right]\right] \tag{15}$$

is

$$Z = \int ...\int dQ_1...dQ_N \exp\left[\frac{i}{2}Q \cdot A \cdot Q + iJ \cdot Q\right] \tag{16}$$

when V(φ) is quadratic, e.g., harmonic oscillator per standard QFT. $A_{ij}$ is the discrete matrix counterpart to the differential operator of Eq. (15) while $J_m$ and $Q_n$ are the discrete vector versions of *J(x)* and *φ(x)*. The discrete action, $\frac{1}{2}Q \cdot A \cdot Q + J \cdot Q$, is considered a functional, which we may write as $\frac{1}{2}|\alpha\rangle\langle\beta| + \langle J|$, of $Q_n$, which we may write as $\langle Q|$ or $|Q\rangle$. Regions in $Q_n$ space where the action is stationary, i.e., invariant/symmetric, contribute most prominently to the transition amplitude[3]. Therefore, the functional is constructed so that what one means by trans-temporal objects, space and time, per $\langle J|$ and $|\alpha\rangle\langle\beta|$ respectively, are self-consistently defined and harbor the desired fundamental symmetries (Figure 3). This is of course similar to the modus operandi of theoretical particle physics, the difference being the discrete formalism allows for (requires) the explicit construct of trans-temporal objects in concert with the spacetime metric whereas the spatiotemporally continuous starting point of QFT harbors tacit assumptions/constraints[4].

The solution to Eq. (16) is

$$Z = \left(\frac{(2i\pi)^N}{\det(A)}\right)^{1/2} \exp\left[-\frac{i}{2}J \cdot A^{-1} \cdot J\right] \quad (17).$$

Since $A_{ij}$ has an inverse, it has a non-zero determinant so it's composed of N linearly independent vectors in its N-dimensional, representational vector space. Thus, any vector in this space may be expanded in the set of vectors comprising $A_{ij}$. Specifically, the vector $J_m$, which will be used to represent 'sources' in the experimental set-up, can be expanded in the vectors of $A_{ij}$. In this sense it is clear how relations, represented by $A_{ij}$, can be fundamental to relata, represented by $J_m$. In the case of two coupled harmonic oscillators we have

$$V(q_1, q_2) = \sum_{a,b} \frac{1}{2} k_{ab} q_a q_b = \frac{1}{2} k q_1^2 + \frac{1}{2} k q_2^2 + k_{12} q_1 q_2$$

---

[3] Each possible experimental outcome of a given experiment requires its own "all at once" description yielding its own transition amplitude. For the case of spatially discrete sources, Z is the probability amplitude so it provides a frequency over the possible outcomes via the Born rule.

[4] That one must explicitly construct the trans-temporal objects, space and time of the discrete action suggests there may exist a level of formalism fundamental to the action. Toffoli[57] has proposed that a mathematical tautology resides at this most fundamental level, e.g., "the boundary of a boundary is zero" whence general relativity and electromagnetism[58]. Elsewhere, using discrete graph theory, we propose a self-consistency criterion which is also based on this tautology (quant-ph/0712.2778).

where $k_{11} = k_{22} = k$ and $k_{12} = k_{21}$, so our Lagrangian is

$$L = \frac{1}{2}m\dot{q}_1^2 + \frac{1}{2}m\dot{q}_2^2 - \frac{1}{2}kq_1^2 - \frac{1}{2}kq_2^2 - k_{12}q_1q_2$$

and the spatially and temporally discrete version of $A_{ij}$ in Eq. (16) would be

$$A = -\begin{pmatrix} \frac{m}{\Delta t}+k\Delta t & \frac{-2m}{\Delta t} & \frac{m}{\Delta t} & 0 & \ldots & 0 & k_{12}\Delta t & 0 \\ 0 & \frac{m}{\Delta t}+k\Delta t & \frac{-2m}{\Delta t} & \frac{m}{\Delta t} & 0 & \ldots & 0 & k_{12}\Delta t \\ & & & \ddots & & & & \\ k_{12}\Delta t & 0 & \ldots & \frac{m}{\Delta t}+k\Delta t & \frac{-2m}{\Delta t} & \frac{m}{\Delta t} & 0 & \ldots \\ 0 & k_{12}\Delta t & 0 & \ldots & \frac{m}{\Delta t}+k\Delta t & \frac{-2m}{\Delta t} & \frac{m}{\Delta t} & 0 \\ & & & \ddots & & & & \end{pmatrix} \quad (18).$$

The process of temporal identification $Q_n \rightarrow q_i(t)$ may be encoded in the blocks along the diagonal of $A_{ij}$ whereby the spatial division between the $q_i(t)$ would then be encoded in the relevant off-diagonal (interaction) blocks:

$$A = \begin{pmatrix} \begin{array}{|c|c|} \hline \ddots & \\ & q_1(t) \\ & & \ddots \\ \hline \end{array} & \begin{array}{c} q_1(t) \Leftrightarrow q_2(t) \\ \end{array} \\ \begin{array}{c} q_2(t) \Leftrightarrow q_1(t) \\ \end{array} & \begin{array}{|c|c|} \hline \ddots & \\ & q_2(t) \\ & & \ddots \\ \hline \end{array} \end{pmatrix}.$$

The discrete formulation illustrates nicely how NRQM tacitly assumes an *a priori* process of trans-temporal identification, $Q_n \rightarrow q_i(t)$. Indeed, there is no principle which dictates the construct of trans-temporal objects fundamental to the formalism of dynamics in general – these objects are "put in by hand." Thus, RBW suggests the need for a fundamental principle which would explicate the trans-temporal identity employed tacitly in NRQM, QFT and all dynamical theories. Since our starting point does not contain trans-temporal objects, space or time, we have to formalize counterparts to these concepts. Clearly, the process $Q_n \rightarrow q_i(t)$ is an organization of the set $Q_n$ on two levels — there is the split of the set into subsets, one for each 'source', and there is the ordering over each subset. The split represents space (true multiplicity from apparent identity), the ordering represents time (apparent identity from true multiplicity)[5] and the result is objecthood (via relations). Again, the three concepts are inextricably linked in our formalism, thus our suggestion that they be related via a self-consistency criterion (Figure 3).

In the limit of temporal continuity, Eq. (18) dictates we find the inverse of

$$-\begin{pmatrix} m\dfrac{d^2}{dt^2} + k & k_{12} \\ k_{12} & m\dfrac{d^2}{dt^2} + k \end{pmatrix}$$

to obtain Eq. (17) so that

$$-\frac{1}{2} Q \cdot A \cdot Q \rightarrow \int \left( \frac{m}{2} q_1 \ddot{q}_1 + \frac{1}{2} k q_1^2 + \frac{m}{2} q_2 \ddot{q}_2 + \frac{1}{2} k q_2^2 + k_{12} q_1 q_2 \right) dt$$

in our NRQM action. Solving

$$-\begin{pmatrix} m\dfrac{d^2}{dt^2} + k & k_{12} \\ k_{12} & m\dfrac{d^2}{dt^2} + k \end{pmatrix} D_{im}(t-t') = \begin{pmatrix} \delta(t-t') & 0 \\ 0 & \delta(t-t') \end{pmatrix}$$

for $D_{im}(t - t')$ we find

$$D_{im}(t-t') = -\begin{pmatrix} \int \dfrac{d\omega}{2\pi} A(\omega) e^{i\omega(t-t')} & \int \dfrac{d\omega}{2\pi} B(\omega) e^{i\omega(t-t')} \\ \int \dfrac{d\omega}{2\pi} B(\omega) e^{i\omega(t-t')} & \int \dfrac{d\omega}{2\pi} A(\omega) e^{i\omega(t-t')} \end{pmatrix}$$

---

[5] These definitions of space and time follow from a fundamental principle of standard set theory, *multiplicity iff discernibility* (W.M. Stuckey, *Phys. Ess.* **12**, 414-419, (1999)).

with

$$A = \frac{\omega^2 m - k}{k_{12}^2 - (\omega^2 m - k)^2} \quad \text{and} \quad B = \frac{k_{12}}{k_{12}^2 - (\omega^2 m - k)^2}.$$

The NRQM amplitude in this simple case is then given by

$$Z(J) \propto \exp\left[-\frac{i}{\hbar} \iint dt dt' J_1(t) D_{12} J_2(t')\right] = \exp\left[\frac{i}{\hbar} \iiint \frac{dt' dt d\omega}{2\pi} \frac{J_1(t) k_{12} e^{i\omega(t-t')} J_2(t')}{k_{12}^2 - (\omega^2 m - k)^2}\right]$$

having restored $\hbar$, used $D_{12} = D_{21}$ and ignored the "self-interaction" terms $J_1 D_{11} J_1$ and $J_2 D_{22} J_2$. Fourier transforms give

$$Z(j) \propto \exp\left[\frac{i}{\hbar} \int \frac{d\omega}{2\pi} \frac{j_1(\omega)^* k_{12} j_2(\omega)}{(k_{12}^2 - (\omega^2 m - k)^2)}\right] \tag{19}$$

with $J_1(t)$ real.

If we now use this amplitude to analyze the twin-slit experiment, we can compare the result to that of Schrödinger wave mechanics and infer the non-separability of spatial distance therein. There are four $J$'s which must be taken into account when computing the amplitude (Figure 4), so we will use Eq. (19) to link $J_1$ with each of $J_2$ and $J_4$, and $J_3$ with each of $J_2$ and $J_4$, i.e., $J_1 \leftrightarrow J_2 \leftrightarrow J_3$ and $J_1 \leftrightarrow J_4 \leftrightarrow J_3$. In doing so, we ignore the contributions from other pairings, i.e., the exact solution would contain one integrand with $Q_n \rightarrow q_i(t)$, i = 1,2,3,4. Finally, we assume a monochromatic source of the form $j_1(\omega)^* = \Gamma_1 \delta(\omega - \omega_o)$ with $\Gamma_1$ a constant, so the amplitude between $J_1$ and $J_2$ is

$$Z(j) \propto \exp\left[\frac{i}{2\pi\hbar} \frac{\Gamma_1 k_{12} j_2(\omega_o)}{(k_{12}^2 - (\omega_o^2 m - k)^2)}\right]$$

whence we have for the amplitude between $J_1$ and $J_3$ via $J_2$ and $J_4$

$$\psi \propto \exp\left[\frac{i}{2\pi\hbar}(\Gamma_1 d_{12} j_2 + \Gamma_2 d_{23} j_3)\right] + \exp\left[\frac{i}{2\pi\hbar}(\Gamma_1 d_{14} j_4 + \Gamma_4 d_{43} j_3)\right] \tag{20}$$

where

$$d_{im} = \frac{k_{im}}{(k_{im}^2 - (\omega_o^2 m - k)^2)} \tag{21}$$

with $\psi$ the NRQM amplitude. [Z corresponds to the NRQM propagator which yields the functional form of $\psi$ between spatially localized sources, as will be seen below.] With the source equidistance from either slit (or, equivalently, with slits replaced by a pair of coherent laser-excited atoms) the phase $\Gamma_1 d_{12} j_2$ equals the phase $\Gamma_1 d_{14} j_4$, so we have the familiar form

$$\psi \propto \exp\left[\frac{i}{2\pi\hbar}(\Gamma_2 d_{23} j_3)\right] + \exp\left[\frac{i}{2\pi\hbar}(\Gamma_4 d_{43} j_3)\right] \tag{22}.$$

Now we need the corresponding result from Schrödinger wave mechanics with free-particle propagator[59]

$$U(x_2, t; x_1, 0) = \sqrt{\frac{m}{2\pi\hbar i t}} \exp\left[\frac{im(x_2 - x_1)^2}{2\hbar t}\right]$$

for a particle of mass $m$ moving from $x_1$ to $x_2$ in time $t$. This 'exchange' particle has no dynamic counterpart in the formalism used to obtain Eq. (22), but rather is associated with the oscillatory nature of the spatially discrete 'source' (see below). According to our view, this propagator is tacitly imbued "by hand" with notions of trans-temporal objects, space and time per its derivation via the free-particle Lagrangian. In short, the construct of this propagator bypasses explicit, self-consistent construct of trans-temporal objects, space and time thereby ignoring the self-consistency criterion fundamental to the action. The self-inconsistent, tacit assumption of a single particle with two worldlines (a "free-particle propagator" for each slit) is precisely what leads to the "mystery" of the twin-slit experiment[6]. This is avoided in our formalism because $Z$ does not represent the propagation of a particle between 'sources', e.g., $q_i(t) \neq x(t)$ as explained *supra*. Formally, the inconsistent, tacit assumption is reflected in $-\frac{1}{2}Q \cdot A \cdot Q \rightarrow \int \left(\frac{m}{2}\dot{x}^2\right)dt$ where ontologically $m$ (which is *not* the same $m$ that appears in our oscillator potential) is the mass of the 'exchange' particle (i.e., purported dynamical/diachronic entity moving between 'sources' – again, the ontic status of this entity is responsible for the "mystery") and $x(t)$ (which, again, is *not* equal to $q_i(t)$) is obtained by *assuming* a particular spatial metric (this assumption *per se* is not responsible for the "mystery"). Its success in producing an acceptable amplitude when integrating over all paths $x(t)$ in space ('wrong' techniques can produce 'right' answers), serves to deepen the "mystery" because the formalism, which requires interference between different spatial paths, is not consistent with its antecedent ontological assumption, i.e., a single particle taking two paths causing a single click or a 'matter wave' distributed throughout space causing a spatially localized detection event. There is no such self-inconsistency in our approach, because $Z$ is not a "particle propagator" but a 'mathematical machine' which measures the degree of symmetry contained in the "all at once" configuration of trans-temporal objects, space and time represented by $A$ and $J$, as explained *supra*. Thus, this NRQM "mystery" results from an attempt to tell a dynamical story in an adynamical situation. Continuing, we have

$$\psi(x_2,t) = \int U(x_2,t;x',0)\psi(x',0)dx'$$

and we want the amplitude between sources located at $x_1$ and $x_2$, so $\psi(x',0) = \alpha\delta(x'-x_1)$ whence

$$\psi_{12} = \alpha\sqrt{\frac{m}{2\pi\hbar it}}\exp\left[\frac{imx_{12}^2}{2\hbar t}\right] = \alpha\sqrt{\frac{m}{2\pi\hbar it}}\exp\left[\frac{ipx_{12}}{2\hbar}\right]$$

where $x_{12}$ is the spatial distance between sources $J_1$ and $J_2$, $t$ is the interaction time and $p = \frac{mx_{12}}{t}$. Assuming the interaction time is large compared to the 'exchange' particle's characteristic time so that $x_{12}$ is large compared to $\frac{\hbar}{p}$ we have

$$\psi = \psi_{23} + \psi_{43} \propto \exp\left[\frac{ipx_{23}}{2\hbar}\right] + \exp\left[\frac{ipx_{43}}{2\hbar}\right] \qquad (23)$$

as the Schrödinger dynamical counterpart to Eq. (22), whence we infer

$$\frac{p}{2\hbar}x_{ik} = \frac{\Gamma_i d_{ik} j_k}{2\pi\hbar}. \qquad (24)$$

---

[6] Per Feynman, the twin-slit experiment "has in it the heart of quantum mechanics. In reality, it contains the *only* mystery" (R.P. Feynman, R.B. Leighton & M. Sands, *The Feynman Lectures on Physics, Vol. III, Quantum Mechanics* (Addison-Wesley, Reading, 1965), p. 1-1).

Assuming the impulse $j_k$ is proportional to the momentum transfer $p$, we have

$$x_{im} \propto \frac{\Gamma_i k_{im}}{\left(k_{im}^2 - (\omega_o^2 m - k)^2\right)} \qquad (25)$$

relating the spatial separation $x_{im}$ of the trans-temporal objects $J_i$ and $J_m$ to their intrinsic ($m$, $k$, $\omega_o$) and relational ($k_{im}$) 'dynamical' characteristics.

As we stated in section 1, the metric of Eq. (25) provides spatial distance only between interacting ($k_{im} \neq 0$) trans-temporal objects, in stark contrast to the metric field of relativity theory which takes on values at each point of the differentiable spacetime manifold, even in regions where the stress-energy tensor is zero. And, as is clear from our presentation, there is no 'exchange' particle or wave (of momentum $p$ or otherwise) moving 'through space' from the source to the detector to 'cause' a detection event. Thus, we have a formal counterpart to our heuristic graphical illustration whereby there is no concept of spatial distance in spacetime regions where the stress-energy tensor vanishes.

**3. RESOLVING THE CONCEPTUAL PROBLEMS OF NRQM**

Before we use RBW to address the conceptual problems of NRQM, we pause to enumerate the RBW ontology and methodology.

1. We may view each piece of equipment in an experimental set-up as resulting from a large number of spatiotemporally dense relations, so low-intensity sources and high-sensitivity detectors must be used to probe the rarified realm of NRQM (Figure 3).
2. A "detector click" or "detector event" is a subset of the detector that also results from a large number of spatiotemporally dense relations; we infer the existence of a rarified set of relations between the source and the detector at the beginning of the click's worldline.
3. It is this inferred, rarified set of relations for which we compute the transition amplitude.
4. A transition amplitude must be computed for each of all possible click locations (experimental outcomes) and this calculation must include (tacitly if not explicitly) all relevant information concerning the spacetime relationships (e.g., distances and angles) and property-defining relations (e.g., degree of reflectivity) for the experimental equipment.
5. The relative probability of any particular experimental outcome can then be determined via the transition amplitude, which is the probability amplitude of NRQM for spatially discrete sources.

*3.1 The Measurement Problem.* According to the account developed here, we offer a deflation of the measurement problem with a novel form of a "statistical interpretation."

The fundamental difference between our version of this view and the usual understanding of it is the following: whereas on the usual view the state description refers to an "ensemble" which is an ideal collection of similarly prepared quantum particles, "ensemble" according to our view is just an ideal collection of spacetime regions $S_i$ "prepared" with the same spatiotemporal boundary conditions per the experimental configuration itself. The union of the click events in each $S_i$, as $i \to \infty$, produces the characteristic Born distribution. Accordingly, probability in RBW is interpreted per relative frequencies.

On our view, the wavefunction description of a quantum system can be interpreted statistically because we now understand that, as far as measurement outcomes are concerned, the Born distribution has a basis in the spacetime symmetries of the experimental configuration. Each "click," which some would say corresponds to the impingement of a particle onto a measurement device with probability computed from the wavefunction, corresponds to spacetime relations in the context of the experimental configuration. The measurement problem *exploits* the possibility of extending the wavefunction description from the quantum system to the whole measurement apparatus, whereas the "all at once" description according to RBW *already includes* the apparatus via the spacetime symmetries instantiated by the *entire* experimental configuration. The measurement problem is therefore a non-starter on our view.

Since a trans-temporal object (such as a detector) possesses properties (to include click distributions) according to a spatiotemporally global set of relations (all trans-temporal objects are defined non-separably in "a vast spatiotemporal mosaic"), one could think of RBW as a local hidden-variable theory (such as BCQM) whereby the relations or symmetries provide the "hidden variables." One can construct a *local* hidden-variable theory if one is willing to claim that systems which presumably have not interacted may nevertheless be correlated. Such correlations appear to require some kind of universal conspiracy behind the observed phenomena, hence Peter Lewis[60] calls such theories "*conspiracy theories*." As he says, "the obvious strategy is the one that gives conspiracy theories their name; it involves postulating a vast, hidden mechanism whereby systems that apparently have no common past may nevertheless have interacted." Independence is the assumption that the hidden variables assigned to the particles are independent of the settings of the measuring devices. If Independence is violated, then a local hidden-variable theory (a conspiracy theory) can in principle account for the Bell correlations. But how *could* Independence be violated? The common cause principle tells us that every systematic correlation between events is due to a cause that they share. As a trivial consequence, systems that have not interacted cannot be systematically correlated, and all appearances indicate that the particles and the measuring devices in EPR-Bell phenomena do not interact before the measurement. Lewis[61] suggests three possibilities for violating Independence:

> Hidden-mechanism theories and backwards-causal theories are both strategies for constructing a local hidden-variable theory by violating Independence. The first of these postulates a mechanism that provides a cause in the past to explain the Bell correlations, and the second postulates a cause in the future. But there is a third strategy that is worth exploring here, namely that the common cause principle is *false*—that some correlations simply require no causal explanation.

Lewis calls the third strategy of denying the common cause principle "acausal conspiracy theories;" RBW can be reasonably characterized in this fashion with the spacetime symmetries playing the role of the hidden-variables. However such a characterization is also misleading in that we are not supplementing NRQM in any standard sense, such as modal interpretations *a la* Bohm. We are not claiming that quantum mechanics is incomplete but that the spacetime symmetries and K4 provide a deeper explanation than NRQM as standardly and *dynamically* conceived. At least at this level, there is no deeper explanation for individual outcomes of quantum experiments than that provided by the structure of K4 and the spacetime symmetries underlying each experimental configuration[7]. The measurement problem arises because of the assumption that the *dynamics* are the deepest part of the explanatory story, the very heart of quantum mechanics, an assumption RBW rejects. In short, RBW provides a *kinematic* (pre-dynamical) solution to the measurement problem.

*3.2 Entanglement and Non-locality.* The blockworld description of an experiment includes its outcomes, and it is possible that outcomes are correlated via symmetries included in the definition of the experiment per the action. Again, the description is "all at once" to include outcomes so if these outcomes are correlated per the action, which was constructed to represent a specific subset of reality instantiated (approximately) by the experiment in question, then there is no reason to expect entanglement will respect any kind of common cause principle. As we stated supra, causality/dynamism are not essential in the algorithm for constructing a blockworld description. Although RBW is *fundamentally* adynamical (relata from relations "all at once," rather than relata from relata in a causal or dynamical structure), it does not harbor non-locality in the odious sense of "spooky action at a distance" as in Bohm for example, i.e., there are no space-like worldlines (implied or otherwise) between space-like separated, correlated outcomes. Again, this is where RBW suggests a new approach to fundamental physics because dynamical entities are modeled fundamentally via relations in "a vast spatiotemporal mosaic" instead of via "interacting" dynamical constituents *a la* particle physics[8].

---

[7] Of course, RBW implies a formalism fundamental to NRQM as shown in section 2. This implication sets RBW apart from mere interpretations of NRQM.

      Our account provides a clear description, in terms of relations in a blockworld, of quantum phenomena *that does not suggest the need for a "deeper" causal or dynamical explanation*. If explanation is simply determination, then our view explains the structure of quantum correlations by invoking what can be called *acausal, adynamical global determination relations*. In NRQM, these "all at once" determination relations are given by the spacetime symmetries which underlie a particular experimental set-up. Not objects governed by dynamical laws, but rather acausal relations per the relevant spacetime symmetries do the fundamental explanatory work according to RBW. We can invoke the *entire* spacetime configuration of the experiment so as to predict, and explain, the EPR-Bell correlations. This then is a geometrical, acausal and adynamical account of entanglement.

      In summary, the spacetime symmetries of an NRQM experiment can be used to construct its quantum density operator, such a spacetime (K4) is one for which simultaneity is relative, and events in the detector region(s) evidence rarified relations between spatially discrete sources, which are trans-temporal objects and thus modeled as temporally continuous (recall from section 2 that NRQM obtains in the temporally continuous, spatially discrete limit of the discrete action). To evidence the explanatory power of this interpretation, we use it to resolve a particularly challenging conundrum in NRQM.

---

[8] This means particles physics per QFT is displaced from its fundamental status (Figure 3).

## 4. RESOLVING THE QUANTUM LIAR PARADOX

We now apply the Bohr *et al.* method to a particular experimental set-up. In two recent articles, Elitzur and Dolev try to establish something like the negation of the blockworld view, by arguing for an intrinsic direction of time given by the dynamical laws of quantum theory[62]. They put forward the strong claim that certain experimental set-ups such as the quantum liar experiment (QLE) "entail inconsistent histories" that "undermine the notion of a fixed spacetime within which all events maintain simple causal relations. Rather, it seems that quantum measurement can sometimes 'rewrite' a process's history[63]." In response, they propose a "spacetime dynamics theory[64]." Certainly, if something like this is true, then blockworld is jeopardized. By applying the geometrical interpretation of quantum mechanics to the "quantum liar" case, we will not only show that the blockworld assumption is consistent with such experiments, but that blockworld *a la* our geometric interpretation provides a non-trivial and unique explanation of such experiments.

*4.1 Mach-Zehnder Interferometer & Interaction-Free Measurements.* Since QLE employs interaction-free measurement[65] (IFM), we begin with an explication of IFM. Our treatment of IFM involves a simple Mach-Zehnder interferometer (MZI, Figure 5; BS = beam splitter, M = mirror and D = detector). All photons in this configuration are detected at D1 since the path to D2 is ruled out by destructive interference. This obtains even if the MZI never contains more than one photon in which case each photon "interferes with itself." If we add a detector D3 along either path (Figures 6a and 6b), we can obtain clicks in D2 since the destructive interference between BS2 and D2 has been destroyed by D3. If we introduce detectors along the upper and lower paths between the mirrors and BS2, obviously we do not obtain any detection events at D1 or D2.

To use this MZI for IFM we place an atom with spin X+, say, into one of two boxes according to a Z spin measurement, i.e., finding the atom in the Z+ (or Z-) box means a Z measurement has produced a Z+ (or Z-) result. The boxes are opaque for the atom but transparent for photons in our MZI. Now we place the two boxes in our MZI so that the Z+ box resides in the lower arm of the MZI (Figure 7). If we obtain a click at D2, we know that the lower arm of the MZI was blocked as in Figure 6a, so the atom resides in the Z+ box. However, the photon must have taken the upper path in order to reach D2, so we have measured the Z component of the atom's spin without an interaction. Accordingly, the atom is in the Z+ spin state and subsequent measurements of X spin will yield X+ with a probability of one-half (whereas, we started with a probability of X+ being unity).

*4.2 Quantum Liar Experiment.* QLE leads to the "quantum liar paradox" of Elitzur & Dolev[66] because it presumably instantiates a situation isomorphic to a liar paradox such as the statement: "this sentence has never been written." As Elitzur & Dolev put it, the situation is one in which we have two distinct non-interacting atoms in different wings of the experiment that could only be entangled via the mutual interaction of a single photon. However one atom is found to have blocked the photon's path and thus it could not interact with the other atom via the photon and the other atom should therefore not be entangled with the atom that blocked the photon's path. But, by violating Bell's inequality, its "having blocked the photon" was affected by the measurement of the other atom, hence the paradox. Our explication of the paradox differs slightly in that we describe outcomes via spin measurements explicitly.

We start by exploiting IFM to entangle two atoms in an EPR state, even though the two atoms never interact with each other or the photon responsible for their entanglement[67] [9]. We simply add another atom prepared as the first in boxes Z2+/Z2- and position these boxes so that the Z2- box resides in the upper arm of the MZI (Figure 8). Of course if the atoms are in the Z1+/Z2- states, we have blocked both arms and obtain no clicks in D1 or D2. If the atoms are in Z1-/Z2+ states, we have blocked neither arm and we have an analog to Figure 5 with all clicks in D1. We are not interested in these situations, but rather the situations of Z1+ *or* Z2- as evidenced by a D2 click. Thus, a D2 click entangles the atoms in the EPR state:

$$\frac{1}{\sqrt{2}}\left(|Z+\rangle_1|Z+\rangle_2 + |Z-\rangle_1|Z-\rangle_2\right) \qquad (26)$$

and subsequent spin measurements with orientation of the Stern-Gerlach magnets in $\Re^2$ as shown in Figure 9 will produce correlated results which violate Bell's inequality precisely as illustrated by Mermin's apparatus[69]. This EPR state can also be obtained using *distinct* sources[70] (Figure 10), so a single source is not necessary to entangle the atoms. In either case, subsequent spin measurements on the entangled atoms will produce violations of Bell's inequality.

---

[9] The non-interaction of the photons and atoms is even more strongly suggested in an analogous experiment, where a super-sensitive bomb is placed in on of the arms of the MZI[68].

Suppose we subject the atoms to spin measurements after all D2 clicks and check for correlations thereafter. A D2 click means that one (and only one) of the boxes in an arm of the MZI is acting as a "silent" detector, which establishes a "fact of the matter" as to its Z spin and, therefore, the other atom's Z spin. In all trials for which we chose to measure the Z spin of both atoms this fact is confirmed. But, when we amass the results from all trials (to include those in which we measured Γ and/or Δ spins) and check for correlations we find that Bell's inequality is violated, which indicates the Z component of spin *cannot* be inferred as "a matter of unknown fact" in trials prior to Γ and/or Δ measurements. This is not consistent with the apparent "matter of fact" that a "silent" detector must have existed in one of the MZI arms in order to obtain a D2 click, which entangled the atoms in the first place. To put the point more acutely, Elitzur and Dolev[71] conclude their exposition of the paradox with the observation that

> *The very fact that one atom is positioned in a place that seems to preclude its interaction with the other atom leads to its being affected by that other atom*. This is logically equivalent to the statement: "This sentence has never been written.[10]"

In other words, *there must be a fact of the matter concerning the Z spins in order to produce a state in which certain measurements imply there was no fact of the matter for the Z spins.*

*4.3 Geometrical Account of QLE.* By limiting any account of QLE to a story about the interactions of objects or entities in spacetime (such as the intersection of point-particle-worldlines, or an everywhere-continuous process *connecting* two or more worldlines), it is on the face of it difficult to account for "interaction-free" measurements (since, naively, a necessary condition for an "interaction" is the "intersection of two or more worldlines"). Since the IFM in this experiment "generated" the entanglement, we can invoke the *entire* spacetime configuration of the experiment so as to predict, and explain, the EPR-Bell correlations in QLE.

---

[10] This quote has been slightly modified per correspondence with the authors to correct a publisher's typo. In the original document they go on to point out that "[we] are unaware of any other quantum mechanical experiment that demonstrates such inconsistency."

Accordingly, spatiotemporal relations provide the ontological basis for our geometric interpretation of quantum theory, and on that basis, explanation (*qua* determination) of quantum phenomena can be offered. According to our ontology of relations, the distribution of clicks at the detectors reflects the spatiotemporal relationships between the source, beam splitters, mirrors, and detectors as described by the spacetime symmetry group – spatial translations and reflections in this case. The relevant 2D irreducible representations (irreps) for 1-dimensional translations and reflections are[72]

$$T(a) = \begin{pmatrix} e^{-ika} & 0 \\ 0 & e^{ika} \end{pmatrix} \quad \text{and} \quad S(a) = \begin{pmatrix} 0 & e^{-2ika} \\ e^{2ika} & 0 \end{pmatrix} \quad (27)$$

respectively, in the eigenbasis of T. *These are the fundamental elements of our geometric description of the MZI.* Since, with this ontology of spatiotemporal relations, the matter-geometry dualism has been collapsed, both "object" and "influence" reduce to *spacetime relations*. We can then obtain the *density matrix* for such a system via its spacetime symmetry group per Bohr *et al.* The "entanglement" is understood as correlated outcomes in an "all at once" description of the experiment per the symmetries of the action.

Consider now Figure 5, with the RBW interpretation of quantum mechanics in mind. We must now re-characterize that experimental set-up in our new geometrical language, using the formalism of Bohr *et al.* Let a detection at D1 correspond to the eigenvector $|1\rangle$ of $T(a)$ (associated with eigenvalue $e^{-ika}$) and a detection at D2 correspond to the eigenvector $|2\rangle$ of $T(a)$ (associated with eigenvalue $e^{ika}$). The source-detector combo alone is simply described by the click distribution $|1\rangle$. The effect of introducing BS1 is to change the click distribution per the unitary operator

$$Q(a_o) \equiv \frac{1}{\sqrt{2}}(I - iS(a_o)) \quad (28)$$

where $a_o \equiv \pi/(4k)$. Specifically,

$$Q(a_o) = \frac{1}{\sqrt{2}}\left[\begin{pmatrix} 1 & 0 \\ 0 & 1 \end{pmatrix} - i\begin{pmatrix} 0 & -i \\ i & 0 \end{pmatrix}\right] = \frac{1}{\sqrt{2}}\begin{pmatrix} 1 & -1 \\ 1 & 1 \end{pmatrix} \quad (29)$$

and

$$|\psi\rangle = Q(a_o)|1\rangle = \frac{1}{\sqrt{2}}\begin{pmatrix} 1 & -1 \\ 1 & 1 \end{pmatrix}\begin{pmatrix} 1 \\ 0 \end{pmatrix} = \frac{1}{\sqrt{2}}\begin{pmatrix} 1 \\ 1 \end{pmatrix} \quad (30).$$

This is an eigenstate of the reflection operator, so introducing the mirrors does not change the click distribution. Introduction of the second beam splitter, BS2, changes the distribution of clicks at D1 and D2 per

$$|\psi_{final}\rangle = Q^+(a_o)|\psi\rangle = \frac{1}{\sqrt{2}}\begin{pmatrix} 1 & 1 \\ -1 & 1 \end{pmatrix}\begin{pmatrix} \frac{1}{\sqrt{2}} \\ \frac{1}{\sqrt{2}} \end{pmatrix} = \begin{pmatrix} 1 \\ 0 \end{pmatrix} \quad (31).$$

Note there is no mention of photon interference here. We are simply describing the distribution of events (clicks) in spacetime (spatial projection, rest frame of MZI) using the fundamental ingredients in this type of explanation, i.e., spacetime symmetries (spatial translations and reflections in the MZI, rotations in the case of spin measurements). What it means to "explain" a phenomenon in this context is to provide the distribution of spacetime events per the spacetime symmetries relevant to the experimental configuration.

To complete our geometrical explanation of QLE we simply introduce another detector (D3 as in Figure 6a, say), which changes the MZI description *supra* prior to BS2 in that the distribution of clicks for the configuration is given by

$$|\psi_{final}\rangle = \begin{pmatrix} Q^+(a_o) & & 0 \\ & \ddots & 0 \\ 0 & 0 & 1 \end{pmatrix}\begin{pmatrix} \frac{1}{\sqrt{2}} \\ 0 \\ \frac{1}{\sqrt{2}} \end{pmatrix} = \begin{pmatrix} \frac{1}{\sqrt{2}} & \frac{1}{\sqrt{2}} & 0 \\ -\frac{1}{\sqrt{2}} & \frac{1}{\sqrt{2}} & 0 \\ 0 & 0 & 1 \end{pmatrix}\begin{pmatrix} \frac{1}{\sqrt{2}} \\ 0 \\ \frac{1}{\sqrt{2}} \end{pmatrix} = \begin{pmatrix} \frac{1}{2} \\ -\frac{1}{2} \\ \frac{1}{\sqrt{2}} \end{pmatrix} \quad (32).$$

Again, we need nothing more than $Q^+$, which is a function of the reflection symmetry operator, $S(a)$, to construct this distribution. And for the distribution of clicks for the configuration in Figure 6b

$$|\psi_{final}\rangle = \begin{pmatrix} 1/\sqrt{2} & 1/\sqrt{2} & 0 \\ -1/\sqrt{2} & 1/\sqrt{2} & 0 \\ 0 & 0 & 1 \end{pmatrix} \begin{pmatrix} 0 \\ 1/\sqrt{2} \\ 1/\sqrt{2} \end{pmatrix} = \begin{pmatrix} 1/2 \\ 1/2 \\ 1/\sqrt{2} \end{pmatrix} \quad (33).$$

Of course, spin measurements using the MZI boxes ("spin measurements on the atoms") are viewed as binary outcomes in space (spin ½) with respect to the orientation of the magnetic poles in a Stern-Gerlach device (SG). This is "how the atom was placed in the boxes according to spin." Successive spin measurements are described via rotation, i.e.,

$$|\psi_2\rangle = \begin{pmatrix} \cos\left(\frac{\theta}{2}\right) & -\sin\left(\frac{\theta}{2}\right) \\ \sin\left(\frac{\theta}{2}\right) & \cos\left(\frac{\theta}{2}\right) \end{pmatrix} |\psi_1\rangle$$

where $|\psi_1\rangle$ is created by a source, magnet and detector and $|\psi_2\rangle$ obtains when introducing a second SG measurement at an angle $\theta$ with respect to the first. The three possible orientations for SG measurments in $\mathfrak{R}^2$ considered here and in the Mermin apparatus (initial X+ orientation aside) are shown in Figure 9. As with MZI outcomes, the description of spin measurement is to be understood via the spatiotemporal relationships between source(s) and detector(s) per the experimental arrangement, i.e., there are no "atoms impinging on the detectors" behind the SG magnets per their spins. There are just sources, detectors and magnets whose relative orientations in space provide the computation of probabilities for event (click) distributions.

This constitutes an acausal and *adynamical* characterization and explanation of entanglement. According to our view, the *structure of correlations* evidenced by QLE is *determined by* the spacetime relations instantiated by the experiment, understood as a spatiotemporal whole (blockworld). This determination is obtained by *describing* the experimental arrangement from beginning to end (including outcomes) via an action which contains the spatiotemporal symmetry structure relevant to the experimental arrangement and is constructed from self-consistently defined trans-temporal objects, space and time. Since

(i) the explanation lies in the spacetime relations evidenced by (inferred from) detector events,

(ii) the distribution of detector events follows from an "all at once" description of the experimental set-up via its spatially discrete action,

(iii) the action is obtained by a self-consistent definition of trans-temporal objects, space and time,

(iv) the self-consistent construct of the action instantiates the relevant, fundamental symmetries characterizing the experiment and

(v) the ontological structural realism of RBW collapses the matter-geometry dualism,

our geometrical quantum mechanics provides for an *acausal*, *global* and *adynamical* understanding of NRQM phenomena.

*4.4 QLE and Blockworld.* Our analysis of QLE shows the explanatory necessity of the reality of all events—in this case the reality of all phases (past, present and future) of the QLE experiment. We can provide an illustrative, though qualitative, summary by dividing the QLE into three spatiotemporal phases, as depicted in Figures 11 – 13. In the first phase the boxes Z1+, Z1-, Z2+, and Z2- are prepared – without such preparation the MZI is unaffected by their presence. In a sense, the boxes are being prepared as detectors since they have the potential to respond to the source ("atom absorbs the photon" in the language of dynamism). The second phase is to place the four boxes in the MZI per Figure 8 and obtain a D1 or D2 click (null results are discarded). The third phase is to remove the four boxes and do spin measurements. The entire process is repeated many times with all possible $\Gamma$, $\Delta$ and Z spin measurements conducted randomly in phase 3. As a result, we note that correlations in the spin outcomes after D2 clicks violate Bell's inequality.

We are not describing "photons" moving through the MZI or "atoms" whose spin-states are being measured. According to our ontology, clicks are evidence not of an impinging particle-in-motion, but of rarified *spacetime relations* which are a subset of the dense set comprising the equipment of the experiment. If a Z measurement is made on either pair of boxes in phase 3, an inference can be made *a posteriori* as to which box acted as a "silent" detector in phase 2. If $\Gamma$ and/or $\Delta$ measurements are done on each pair

(Figure 11), then there is *no fact of the matter* concerning the detector status of the original boxes (boxes had to be recombined to make Γ and/or Δ measurements). This is not simply a function of ignorance because if it was possible to identify the "silent" detectors before the Γ and/or Δ measurements were made, the Bell assumptions would be met and the resulting spin measurements would satisfy the Bell inequality. Therefore, *that none of the four boxes can be identified as a detector in phase 2 without a Z measurement in phase 3 is an ontological, not epistemological, fact* and points to the necessity of an "all at once" explanation.

Notice that what obtains in phase 3 "determines" what obtains in phase 2, so we have a true "delayed-choice" experiment. For example, suppose box Z2- is probed in phase 3 (Z measurement) and an event is registered (an "atom resides therein," Figure 12). Then, the Z2- and Z1- boxes are understood during phase 3 to be detectors in phase 2. However, nothing in the blockworld has "changed" – the beings in phase 2 have not "become aware" of which boxes are detectors. Neither has anything about the boxes in phase 2 "changed." According to our view, the various possible spatiotemporal distributions of events are each determined by NRQM *as a whole throughout space and time*.

To further illustrate the blockworld nature of the correlations, suppose we make spin measurements after a D1 click. Figure 13 shows a spatiotemporal configuration of facts in phases 1, 2 and 3 consistent with a D1 click:

>Phase 1: No prep
>Phase 2: Boxes are not detectors, D1 click
>Phase 3: Γ2 measurement, Δ1 measurement, No outcomes.

One can find correlated spatiotemporal facts by starting in any of the three phases:
Starting with phase 3, "No outcomes" → "No prep" in phase 1 and "Boxes are not detectors" and "D1 click" in phase 2. If you insisted on talking dynamically, you could say that the "No outcomes" result of phase 3 *determined* the "Boxes are not detectors" result of phase 2.
Starting with phase 2, "Boxes are not detectors" → "D1 click" in phase 2, "No prep" in phase 1 and "No outcomes" in phase 3.

Starting with phase 1, "No prep" → "No outcomes" in phase 3 and "Boxes are not detectors" and "D1 click" in phase 2.

One can chart implications from phase 1 to phase 3 then back to phase 2, since the order in which we chart implications in a spacetime diagram is meaningless (meta-temporal) to the blockworld inhabitants. In point of fact the collective characteristics in all three phases of QLE are acausally and globally (without attention to any common cause principle) determined by the spacetime symmetries of the experimental set-up; hence, the explanatory necessity of the blockworld. What *determines* the outcomes in QLE is not given in terms of influences or causes. In this way we resolve the quantum liar paradox with RBW by showing how "the paradox" is not only *consistent* with a blockworld structure, but actually strongly suggests an adynamical approach such as ours over interpretations involving dynamical entities and their histories. It is the *spatiotemporal configuration of QLE as a spacetime whole and its spacetime symmetries* that determine the outcomes and not *constructive (*a la *Einstein) entities with dynamical histories*.

## 5. CONCLUSION

According to our Relational Blockworld interpretation of non-relativistic quantum mechanics, one can do justice to the non-commutative structure of NRQM without being a realist about Hilbert space. The trick is to understand that the spacetime of NRQM is a non-separable, relational blockworld that respects locality per SR. Accordingly, one should not think of this spacetime as an empty vessel waiting to be imbued with worldlines and stress-energy because, per the fundamental self-consistency criterion, the concepts of time and space only have meaning in the context of trans-temporal objects, and vice-versa. While clicks in detectors are perfectly classical events, the clicks are not evidence of *constructive* quantum entities such as particles with worldlines. Rather, the clicks are manifestations of the relations composing elements of the experimental configuration as illustrated, for example, by the way RBW parses the quantum liar experiment via the irreps of spatial translations and reflections. This spacetime respects relativistic locality in that there are no faster-than-light "influences" or "productive" causes between space-like separated events, but it does harbor "all at once" geometric "correlations" outside the lightcone as determined acausally, adynamically and globally by the spacetime symmetries. Once again, such acausal and adynamical global determination relations do not respect any common cause principle. This fact should not bother anyone who has truly transcended the idea that the dynamical or causal perspective is the most fundamental one.

In short, unlike Rovelli's or Mermin's relationalist accounts of quantum mechanics which are still *dynamical* in nature, RBW employs the *spatiotemporal* relations via symmetries of the entire (past, present and future) experimental configuration and is thus fundamentally *kinematical*. And unlike other BW inspired accounts of quantum mechanics such as BCQM, RBW is truly acausal, adynamical and atemporal. As well, unlike other relational accounts, to use Einstein's language RBW characterized as a form of ontological structural realism is a complete break with the explanatory fundamentality of *constructive* (to use Einstein's term) and dynamical explanations.

While this interpretation of NRQM is strongly supported by the work of Kaiser, Anandan, Bohr, Ulfbeck, and Mottelson (referenced extensively herein), we are only now researching its implied adynamical, acausal ontology, whereby relations are fundamental to relata, at the level fundamental to NRQM via a spatiotemporally discrete action. Even though the formalism is incomplete, we have enough to speculate on its consequences for quantum gravity (QG). As with G4 and M4, the spacetime of general relativity (GR4) is an approximation which holds only in the large-order limit of spatiotemporally dense sets of relations. Therefore, we expect the GR4 approximation to break down in the realm of rarefied relations between two or more spatiotemporally dense sets of relations (each dense set requiring a metric per GR), e.g., the exchange of 'entangled particles' between stars in different galaxies[11]. In such cases, the everywhere separable metric of GR4 (providing continuously a distance in the empty space between galaxies) must be superceded by a discrete, non-separable metric *a la* that for spatial distance in Eq. 25. This implies the classical spacetime metric (for dense relations) is only a statistical approximation. Since spatiotemporal relationships can only be self-consistently defined in the context of trans-temporal objects, it must be the case that the stress-energy tensor is also a statistical approximation. Classically, the stress-energy tensor can be obtained by the variation of the matter-energy Lagrangian with respect to the metric, so Einstein's equations are probably a classical limit to the proposed self-consistency criterion for space, time and trans-temporal objects of our spatiotemporally discrete formalism (Figure 3).

---

[11] This is distinct from the regime typically understood for QG, i.e., regions where *large* energy densities give rise to GR singularities.

QG so obtained would not be viewed as a fundamental theory of physics. Rather, QG in this context is just another limiting case of the (relevant) discrete action. Since the discrete action is to be obtained via a self-consistent definition of space, time and trans-temporal objects, there is no "problem of time" and we automatically have a background independent formulation. Thus, RBW produces a new direction for QG research which stems from "two things: the foundations of quantum mechanics and the nature of time," as predicted by Smolin.

**ACKNOWLEDGEMENT**

We are very grateful for extensive, insightful and detailed anonymous referee comments on previous versions of the manuscript.

**Figure 1**

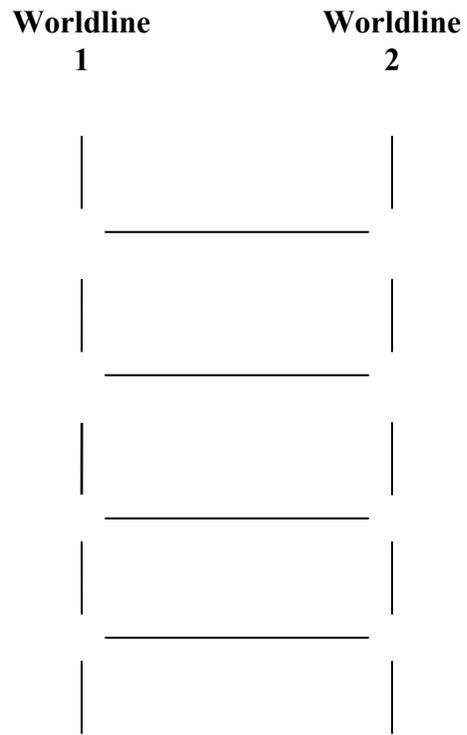

**Figure 2**

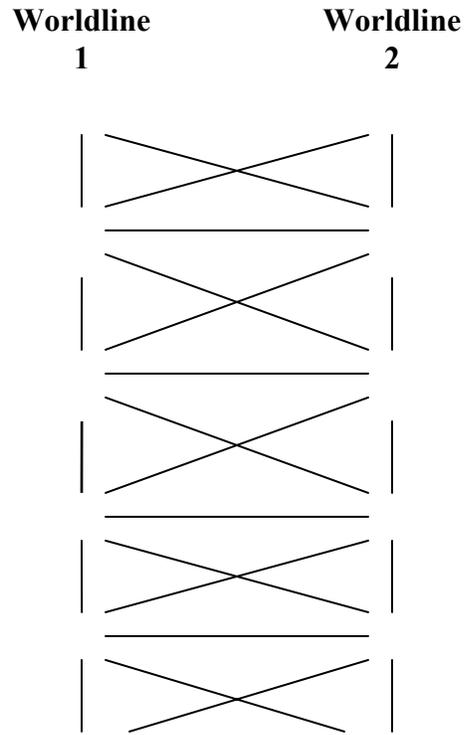

**Figure 3**

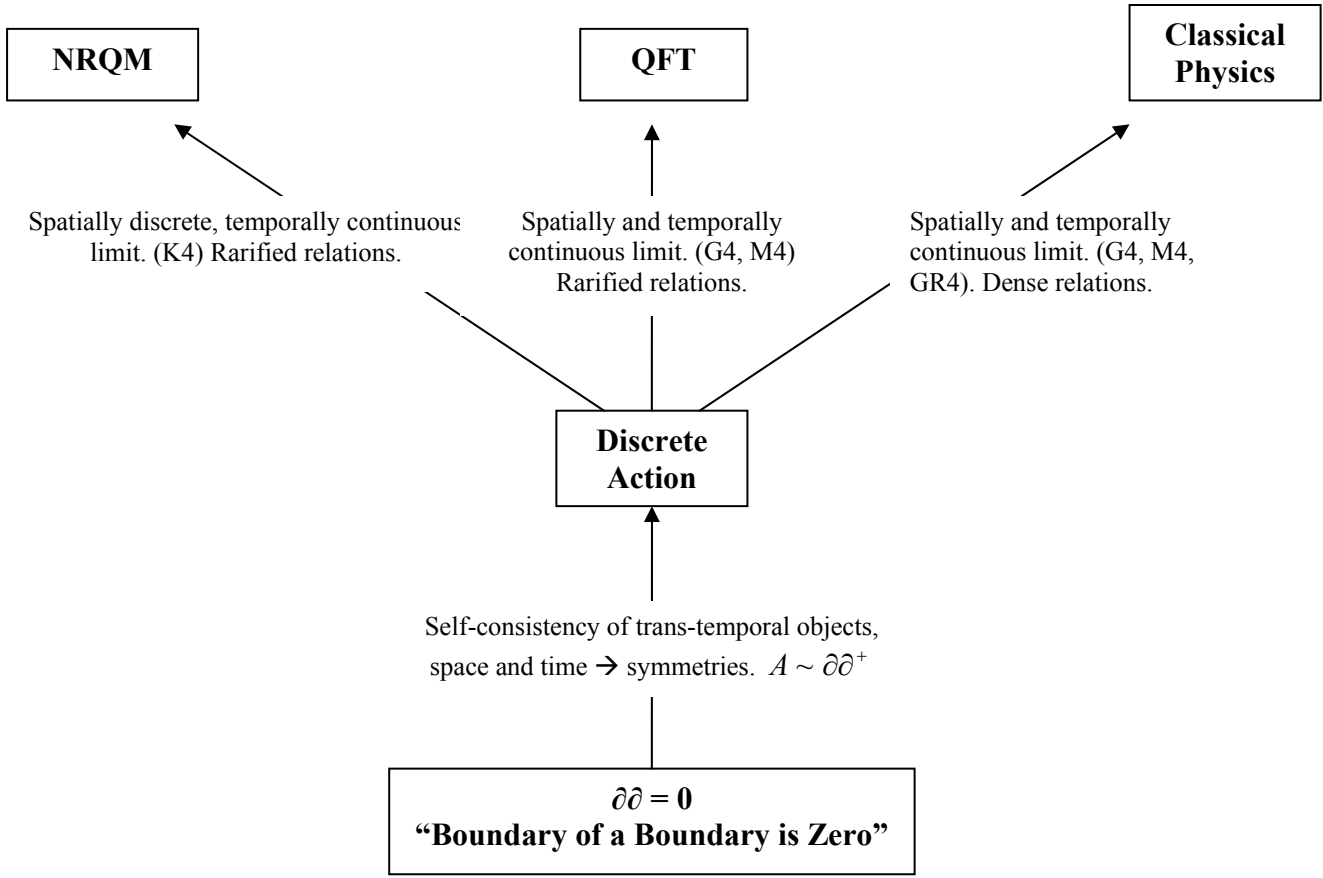

**Figure 4**

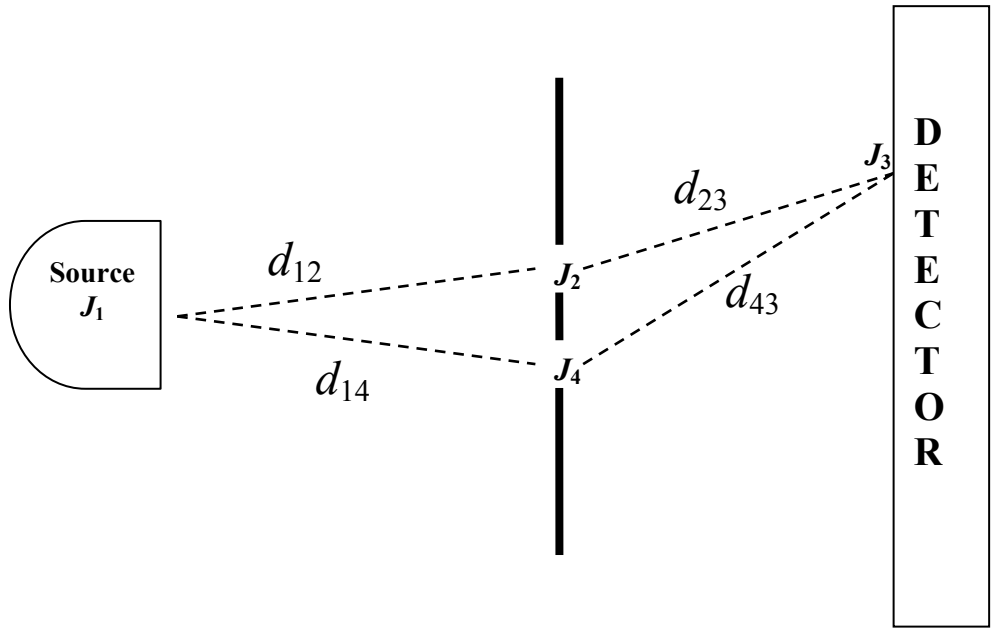

**Figure 5**

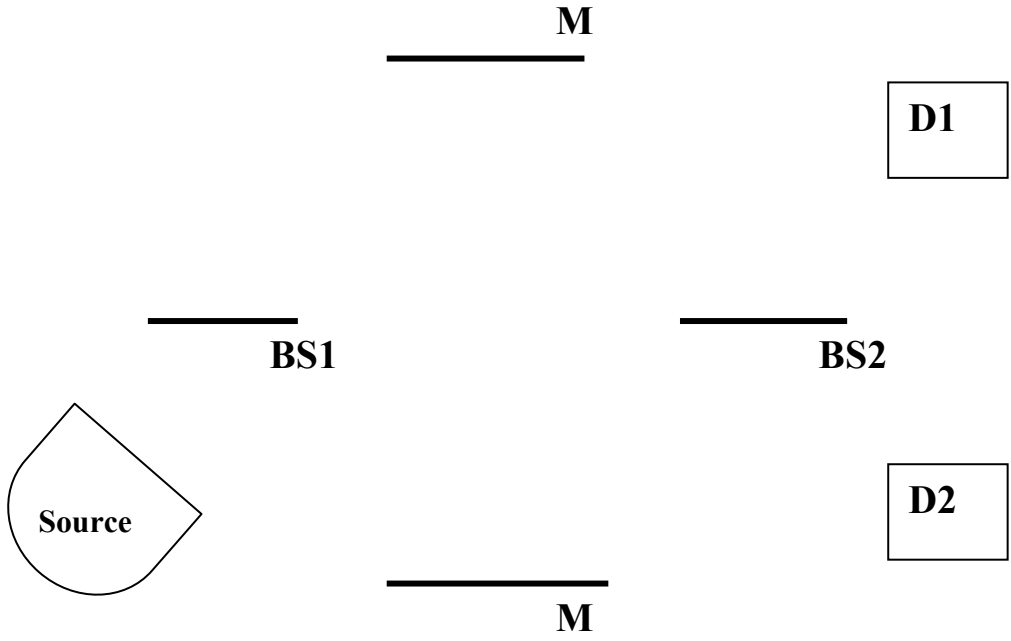

**Figure 6a**

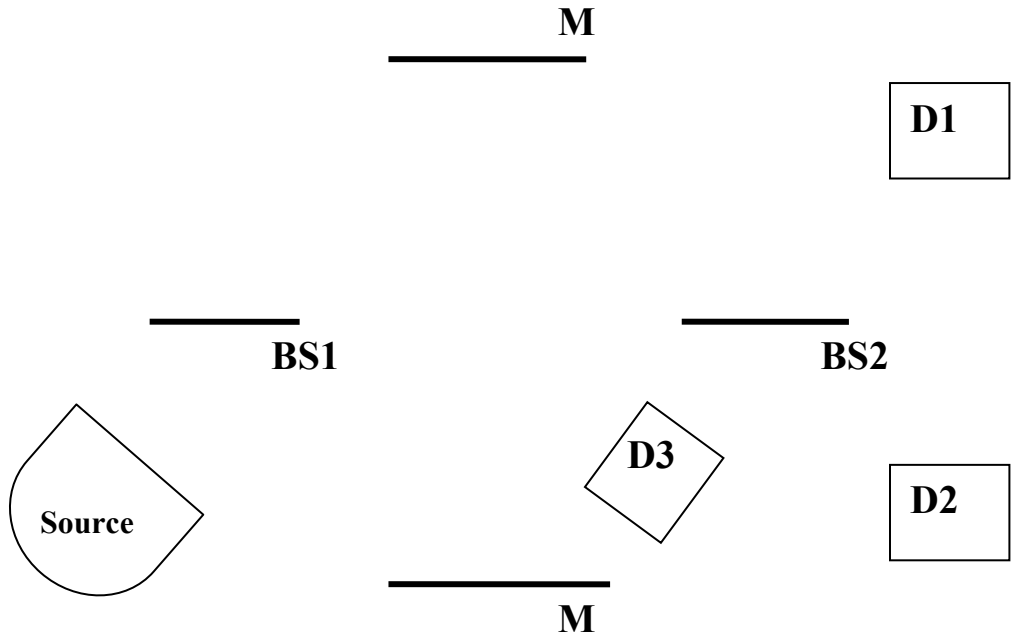

**Figure 6b**

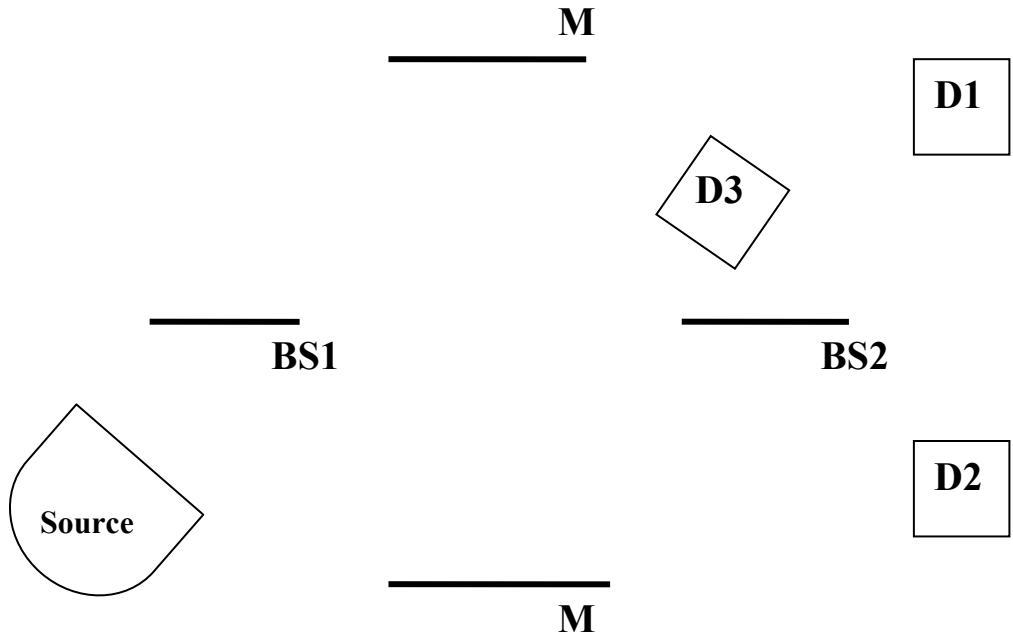

**Figure 7**

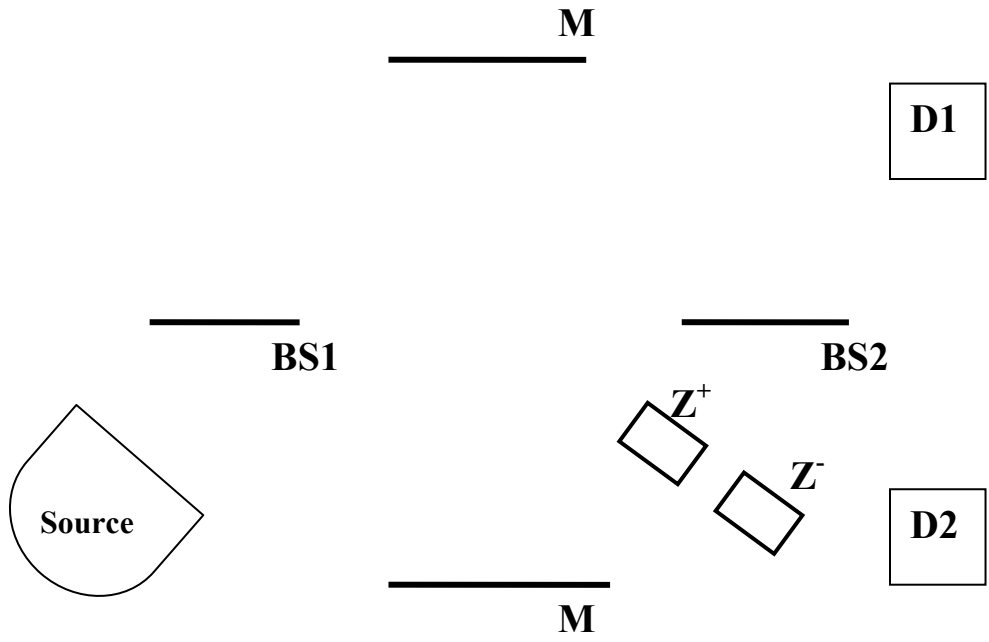

**Figure 8**

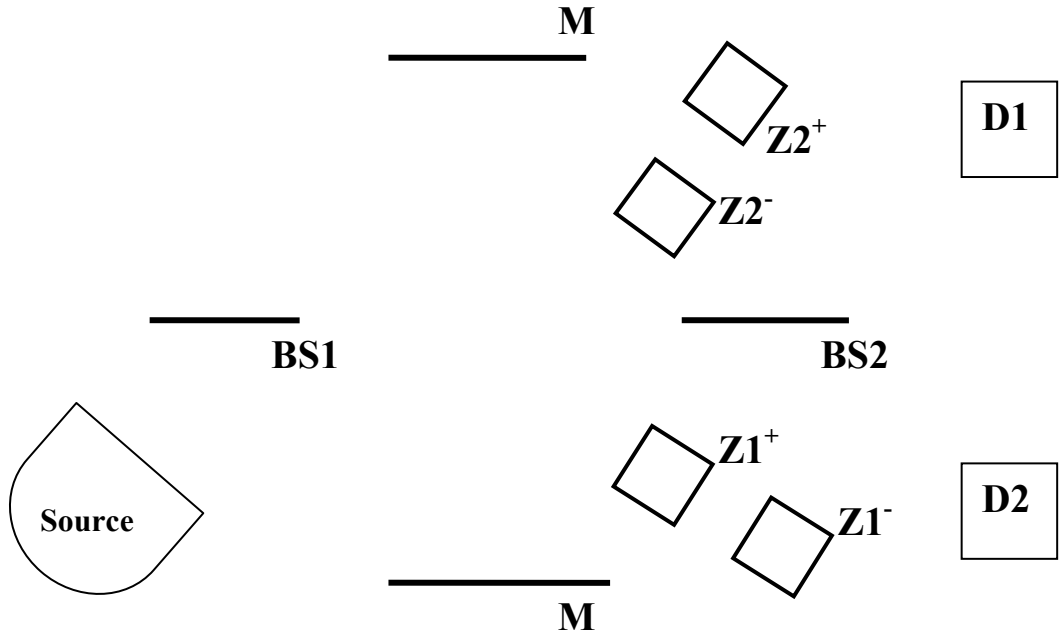

**Figure 9**

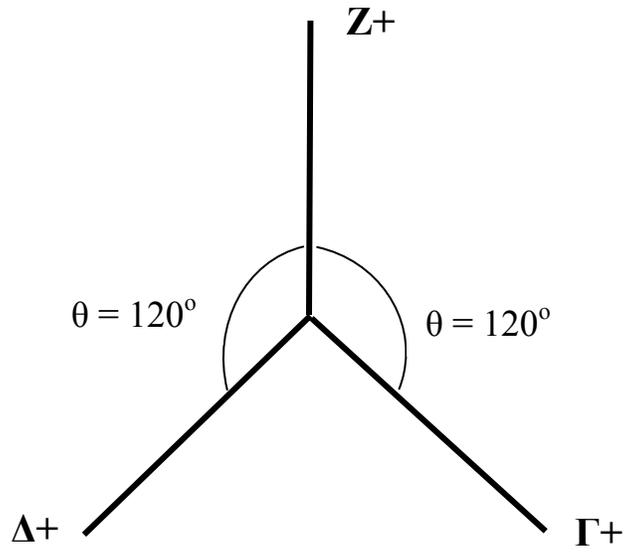

**Figure 10**

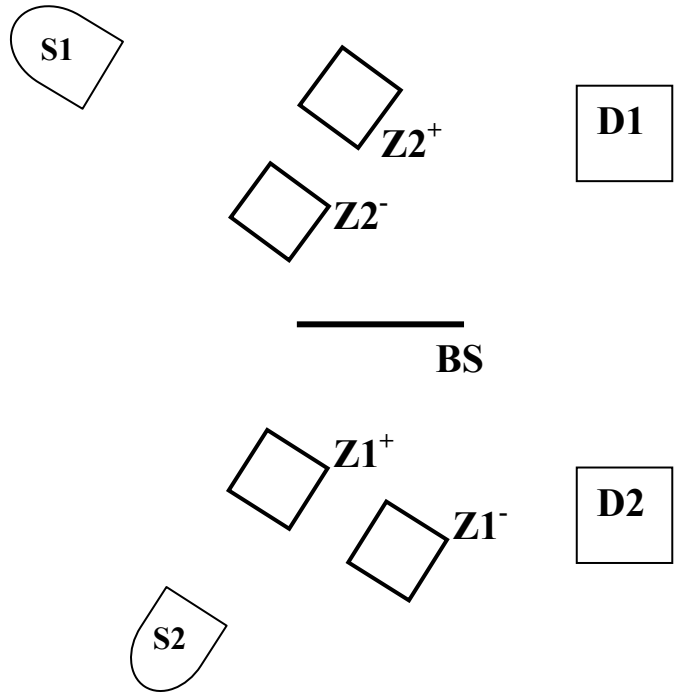

**Figure 11**

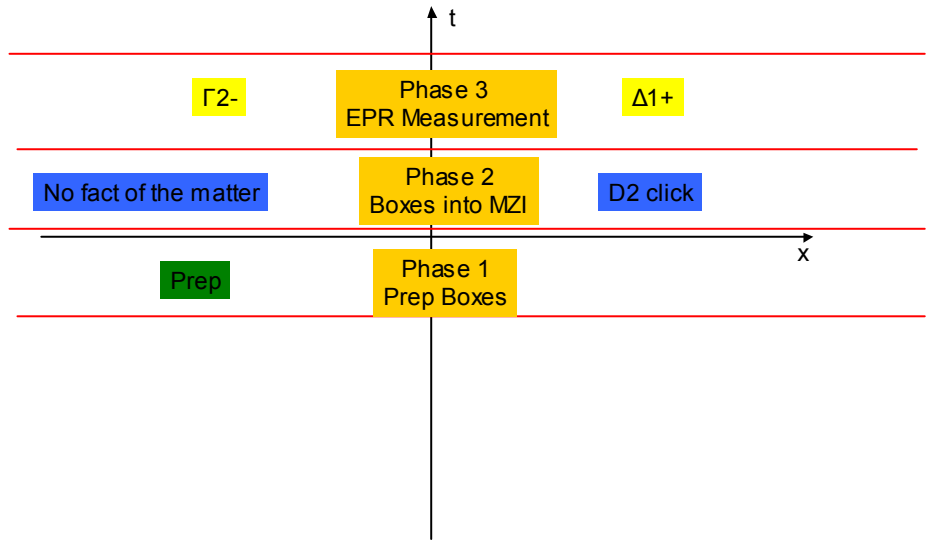

**Figure 12**

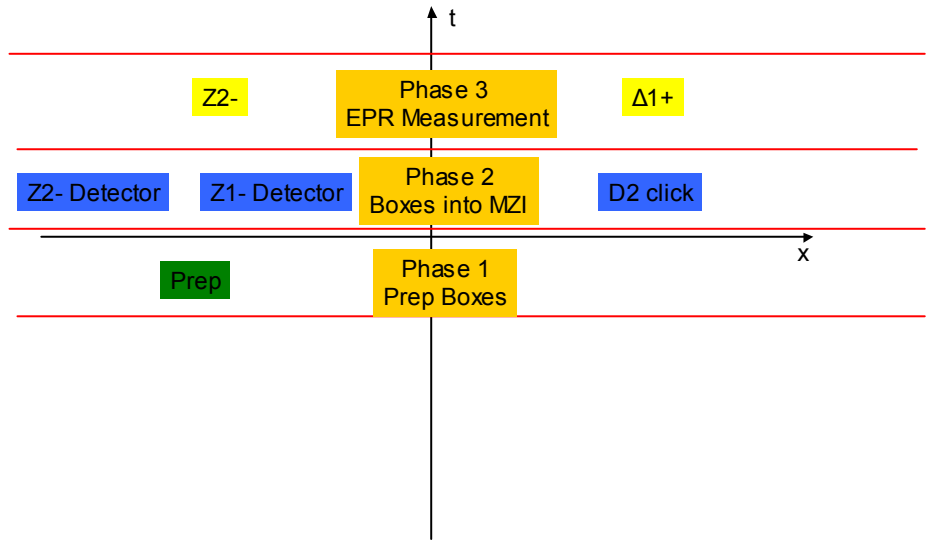

**Figure 13**

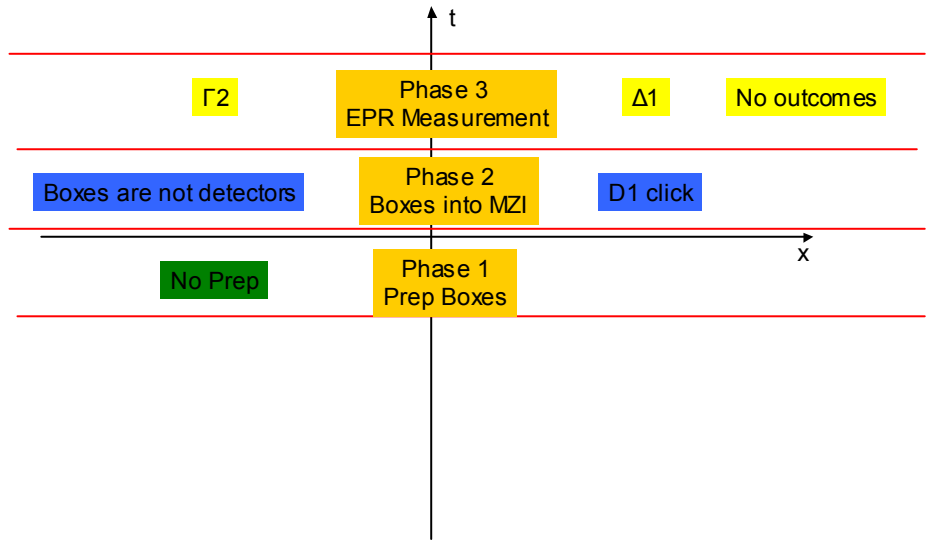